\documentclass[sigplan,nonacm]{acmart}
\usepackage{qcircuit}
\usepackage{amsmath,amsfonts,multirow, array,graphicx,mathtools,bm,times,tcolorbox,relsize,booktabs, subfigure}
\usepackage{stfloats}
\usepackage{algorithmic}

\newcommand{\nc}{\newcommand}
\nc{\bra}[1]{\langle#1|}
\nc{\ket}[1]{|#1\rangle}
\nc{\ketbra}[2]{|#1\rangle\!\langle#2|}
\nc{\braket}[2]{\langle#1|#2\rangle}

\usepackage[ruled, vlined, linesnumbered]{algorithm2e}

\begin{document}
\title{S-SYNC: Shuttle and Swap Co-Optimization in \\ Quantum Charge-Coupled Devices}
\author{Chenghong Zhu}
\authornote{Co-first authors.}
\affiliation{%
  \institution{Thrust of Artificial Intelligence, Information Hub, \\ The Hong Kong University of Science and Technology (Guangzhou)}
  \state{Guangdong}
  \country{China}
  \postcode{511453}
}

\author{Xian Wu}
\authornotemark[1]
\affiliation{%
  \institution{Thrust of Artificial Intelligence, Information Hub, \\ The Hong Kong University of Science and Technology (Guangzhou)}
  \state{Guangdong}
  \country{China}
  \postcode{511453}
}

\author{Jingbo Wang}
\authornote{Co-corresponding authors.}
\affiliation{\institution{Beijing Academy of Quantum Information Sciences}
\city{Beijing}
\country{China}
}
\email{wangjb@baqis.ac.cn}

\author{Xin Wang}
\authornotemark[2]
\affiliation{%
  \institution{Thrust of Artificial Intelligence, Information Hub, \\ The Hong Kong University of Science and Technology (Guangzhou)}
  \state{Guangdong}
  \country{China}
  \postcode{511453}
}
\email{felixxinwang@hkust-gz.edu.cn}

\begin{abstract}
    The Quantum Charge-Coupled Device (QCCD) architecture is a modular design to expand trapped-ion quantum computer that relies on the coherent shuttling of qubits across an array of segmented electrodes. Leveraging trapped ions for their long coherence times and high-fidelity quantum operations, QCCD technology represents a significant advancement toward practical, large-scale quantum processors. However, shuttling increases thermal motion and consistently necessitates qubit swaps, significantly extend execution time and negatively affect application success rates. In this paper, we introduce S-SYNC -- a compiler designed to co-optimize the number of shuttling and swapping operations. S-SYNC exploits the unique properties of QCCD and incorporates generic SWAP operations to efficiently manage shuttle and SWAP counts simultaneously. Building on the static topology formulation of QCCD, we develop scheduling heuristics to enhance overall performance. Our evaluations demonstrate that our approach reduces the shuttling number by 3.69x on average and improves the success rate of quantum applications by 1.73x on average. Moreover, we apply S-SYNC to gain insights into executing applications across various QCCD topologies and to compare the trade-offs between different initial mapping methods.
\end{abstract}

\maketitle 

\section{Introduction}
Quantum computing represents a highly advanced field at the forefront of modern technology, employing the principles of quantum mechanics to enhance information processing capabilities, and the prospective applications of quantum computing are extensive and diverse, ranging from drug discovery~\cite{cao2018potential} to addressing intricate financial issues~\cite{orus2019quantum} and even extending to the realm of artificial intelligence~\cite{dunjko2018machine}. The viability of quantum computing has been successfully demonstrated on several hardware platforms, like superconducting circuits~\cite{nakamura1999coherent}, 
photonics~\cite{knill2001scheme}, silicon quantum dots~\cite{pla2012single}, Rydberg atoms~\cite{lukin2001dipole}, trapped-ions~\cite{cirac1995quantum}. Presently, the field of quantum computing is evolving from the Noisy Intermediate-Scale Quantum (NISQ) era towards Fault-Tolerant Quantum Computing (FTQC), and technology towards FTQC called quantum error correction (QEC)~\cite{devitt2013quantum} is essential. Among these, trapped ions and Rydberg atoms have shown their promise to demonstrate QEC~\cite{postler2022demonstration, bluvstein2023logical, da2024demonstration}. These advancements provide a feasible pathway for the realization of universal quantum computing.

Trapped-ion technology has firmly established itself as a preeminent platform, marked by rapid technological advancements. A breakthrough in this area is the significant extension of qubit coherence times, surpassing the hour~\cite{wang2021single}. Furthermore, the fidelity of quantum gate operations and readouts has reached the critical threshold required for QEC~\cite{harty2014high, gaebler2016high}. Notably, many enhancements control techniques are being reported in trapped-ion platforms, such as amplitude, phase, and frequency modulation \cite{roos2008ion,milne2020phase,leung2018robust}, making trapped-ion a widely used demonstration platform for quantum computing.

Full connectivity among qubits in trapped-ion boasts significant advantages, greatly facilitating the execution of complex quantum algorithms that rely on long-range two-qubit gates. However, this full connectivity result from long-range Coulomb interaction also presents challenges, particularly in terms of scalability. As the number of qubits in a single trap increases, phonon frequency crowding significantly complicates quantum operations and makes individual addressing of ions more difficult \cite{landsman2019two}. Furthermore, an increased number of ion qubits necessitates additional laser or microwave control channels, further complicating the scalability of trapped-ion technology.

To scale up trapped-ion devices, researchers have proposed various schemes. One such scheme resembles a Turing tape mechanism, employing a fixed laser interaction zone, while allowing the linear ion crystal to be moved or split as a whole chain~\cite{wu2021tilt}, thus exceeding the number of qubits achievable by fixed laser channels. Another approach considers using higher dimensions to increase the number of qubits. This method abandons the one-dimensional linear confinement, in favor of a two-dimensional ion crystal, leveraging localized phonon modes to achieve quantum entanglement~\cite{bruzewicz2016scalable}. Additionally, a scheme involving the connection of multiple trapped-ion chips via optical fibers has been proposed, where the expansion of qubits is achieved through the exchange of photons between ions and fibers photons~\cite{monroe2013scaling}.

Among these, the Quantum Charge-Coupled Device (QCCD) \cite{wineland1998experimental,kielpinski2002architecture} has emerged as a pivotal approach, witnessing rapid development in recent years~\cite{wan2019quantum, kaufmann2017scalable, lekitsch2017blueprint, pino2021demonstration, moses2023race, malinowski2023wire, murali2022toward}.
\cite{da2024demonstration} also highlights the advantage of the QCCD module's design for QEC on a racetrack-style trapped-ion QCCD chip. QCCD entails segmenting a multi-dimensional trapped-ion chip into various zones designated for ion operations, storage, and cooling. The movement of ions among these zones is meticulously orchestrated using micro-electrodes, thereby enabling the expansion and precise control of ion qubits. This technique has shown immense promise in the near-term realization of scalable quantum computing. However, ion movement can increase thermal motion, introducing errors during the application of multi-qubit gates. This presents a significant challenge in efficiently managing movements between scalable trapped-ion devices.

Ion movement in QCCD devices serves a role analogous to SWAP insertion in superconducting devices, both introducing additional costs to quantum algorithm execution due to devices' respective physical constraints.
However, a fundamental difference lies in the formulation of these operations. The standard topology graph for superconducting device cannot be directly applied to QCCD devices, as ion movement dynamically alters the topology, complicating the scheduling of these operations. Additionally, we observe that ion movement operations are frequently accompanied by SWAP operations to reorder the ion chain, which can further impact the success rate of quantum applications on QCCD devices.

To address this, we propose S-SYNC, a scheduling framework that coordinates ion movement within traps while minimizing the SWAP gates required for quantum algorithm execution. Our benchmarking across various applications demonstrates significant performance improvements.

The main contribution are summarized as following,
\begin{itemize}
    \item \textbf{Generic Swap Formulation.} 
    We model the topology of QCCD as a weighted connectivity graph, with weights representing the additional costs associated with shuttling and swapping. This model addresses previous issues in QCCD formulations where the topology graph would change after each shuttling operation. Based on it, we then introduce a unified operation, termed `generic swap', which combines the functions of SWAP and shuttle within QCCD devices.
    \item \textbf{Scalable Algorithm Design.} 
    Leverage the static topology formulation of QCCD, we introduce a heuristic search algorithm based on generic-swap operations. This approach allows us to minimize shuttling costs while simultaneously optimizing SWAP gate usage, achieving co-optimization of shuttle and SWAP operations. Additionally, we propose several initial qubit mapping methods specifically tailored for QCCD devices.
    \item \textbf{Enhanced Performance.} We compare our algorithm with previous QCCD compiler support and observe significant enhancements in shuttle number and success rate across different quantum applications. We demonstrate that our approach reduces the shuttling number by 3.69x and improves the success rate of quantum applications by 1.73x on average. We further benchmark our algorithm on different topologies and identify trade-off between different initial mapping methods.
\end{itemize}

\begin{figure}[t]
    \centering
    \includegraphics[width=1\linewidth]{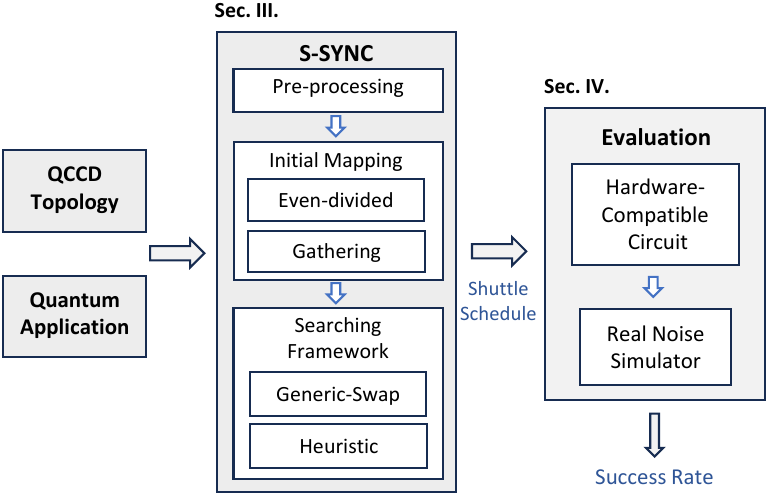}
    \caption{Overview of this work.  We propose S-SYNC with equipment of different initial mapping and searching framework for optimizing the success rate of running quantum application in QCCD devices. 
    }
    \label{fig:qccd_overview}
\end{figure}

The overview of this work is in Fig.~\ref{fig:qccd_overview}. The rest of this paper is organized as follows.  We introduce the background of quantum computation and trapped-ion hardware in Section.~\ref{sec:background}.  We then propose our algorithm in Section.~\ref{sec:algorithm}. \textcolor{black}{The experimental settings are discussed in Section~\ref{sec:evaluation}, and the evaluation results are presented in Section~\ref{sec:architecture_finding}.}  Related works are summarized in Section.~\ref{sec:related_work} and we conclude the paper in Section.~\ref{sec:conclusion}.

\section{Background}\label{sec:background}

\subsection{Quantum Computation}

\textcolor{black}{A classical bit is the fundamental unit of information, existing in one of two deterministic states `0' and `1'. In quantum computing, the basic unit of quantum information is the qubit, which has two basis states, typically denoted as $\ket{0}$ and $\ket{1}$. Unlike classical bits, a qubit can exist in a superposition of these basis states, expressed as $\ket{\psi} = \alpha\ket{0} + \beta\ket{1}$, where $\alpha,\beta\in \mathbb{C}$ and $|\alpha|^2 + |\beta|^2 = 1$. Quantum gates serve as fundamental operations for manipulating qubit states. A single-qubit gate acts on an individual qubit, while two-qubit gates, such as the controlled-X (CNOT) gate and the M{\o}lmer-S{\o}rensen gate, are used to generate entanglement.} Analogous to the role of universal gates in classical computing, these single and two-qubit gates collectively enable the construction of universal gates for quantum computing systems, allowing for the execution of any quantum operation.

\begin{figure*}[htbp]
    \centering
    \includegraphics[width=1\linewidth]{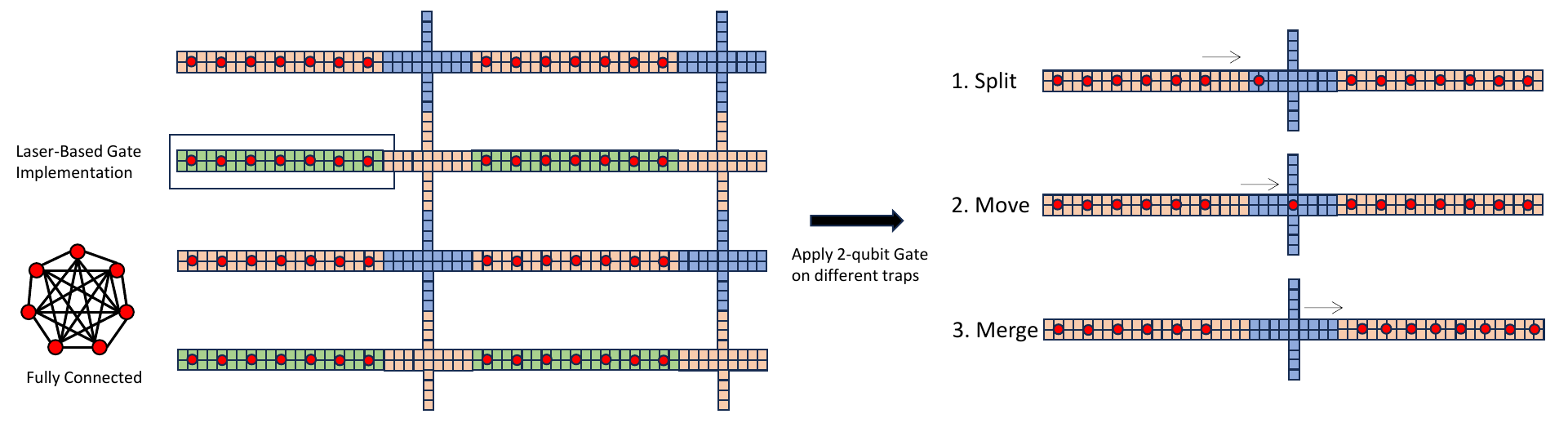}
    \caption{A modular Quantum Charge Coupled Device (QCCD) consists of several traps, each of which is initialized with 7 qubits. These traps are interconnected with shuttle paths. In order to enable the implementation of a two-qubit gate between two traps, ions need to be split from one trap, moved, and then merged into another trap by traversing the shuttle path.
    }
    \label{fig:qccd_whole}
\end{figure*}

\subsection{Trapped-Ion}

Trapped-ion quantum computing employs ions as qubits to execute quantum information operations, this method primarily involves rotating single-qubit gates and MS gates~\cite{haffner2008quantum}. The rotating single-qubit gates achieve precise control over the internal states of ions, while MS gates establish quantum entanglement between two qubits. A significant advantage of trapped-ion quantum computing lies in the high stability of the system's energy levels and the precision of laser modulation techniques, ensuring that both single and two-qubit gates in trapped-ions achieve remarkably high accuracy~\cite{harty2014high, gaebler2016high}. Additionally, the natural non-equidistance of ion internal energy levels facilitates the effective cooling of ions to their ground state using Doppler and sideband cooling techniques, laying a solid foundation for high-fidelity quantum state preparation~\cite{harty2014high}. Fluorescence detection techniques further enable high-fidelity quantum state readouts~\cite{myerson2008high}. 

While small-scale trapped-ion systems have showcased a range of notable benefits, these advantages tend to diminish when attempting to scale up the system. \textcolor{black}{The core issue is that as more qubits are integrated into a trapped-ion system, individually addressing them becomes more challenging, which in turn reduces the precision of quantum gates.} As a result, carrying out quantum computations becomes increasingly difficult, which presents significant hurdles for achieving fault-tolerant quantum computing in the foreseeable future. Consequently, the pursuit of larger qubit counts within a single trapped-ion trap may not be the most effective approach for scaling. This realization has led to a pivot in research efforts towards QCCD architectures, which provide a more promising pathway for the scalable execution of quantum computations.

\subsection{Quantum Charge-Coupled Device}

\textcolor{black}{\textbf{Introduction to QCCD.}} QCCD architecture is distinguished by its modular design, where each structure consists of microstructures dedicated to specific functions like manipulation, storage, and auxiliary zones. These microstructures are seamlessly integrated on a single trapped-ion chip. Each area is capable of independently operating ions, and with the strategic design of micro-electrodes, ions can be transferred between these processing units, enabling efficient quantum information transfer across different units. This design not only enhances the scalability of the system but also reduces the need for global control by allowing local operations, which is crucial for building practical large-scale quantum computers. Moreover, the QCCD architecture excels in maintaining quantum state stability and executing efficient quantum gate operations.

Following the proposal of the QCCD modular expansion protocol by Wineland~\cite{wineland1998experimental,kielpinski2002architecture}, efforts have been initiated to fabricate trapped-ion quantum chips based on the QCCD architecture. As a critical avenue in the scalable exploration of trapped-ion quantum computing, the field, despite facing challenges related to material, optical fibers integration and electrode design, has consistently drawn interest for intellectual contributions. This domain represents a significant frontier in quantum computing, inviting ongoing efforts to overcome its technical obstacles~\cite{monroe2013scaling,ivory2021integrated,brown2021materials}.

Advances in fabricating trapped-ion chips made it possible to experimentally demonstrate QCCD on a small scale. \textcolor{black}{Quantinuum} emerged as a pioneer in this realm, showcasing a linear QCCD chip~\cite{pino2021demonstration}. Despite various environmental noise factors, they achieved a 64 quantum volume. Later then, they demonstrated a two dimensional race track-style QCCD trapped-ion chip. This periodic QCCD chip significantly enhanced the efficiency of quantum circuit execution, achieving remarkably low levels of quantum gate distortion and readout errors, and elevating the quantum volume to $2^{16}$ \cite{moses2023race}. These advancements underscore the potential of QCCD structures as a viable pathway for large-scale, practical implementation of trapped-ion chips.

\textcolor{black}{\textbf{Operations in QCCD.}} Since QCCD is a modular design, it is very critical to make ions connect between different modules. The key operations are `split', `move' and `merge', as shown in Fig.~\ref{fig:qccd_whole}.
In detail, `split' refers to the spatial disentanglement of one ion from a ion crystal, effectively isolating it from interaction with the others. Conversely, `merge' is the reverse process, where a lone ion is integrated into an ion crystal, becoming part of it, and thus capable of engaging in two-qubit gate operations with other ions in the chain.

The `linear shuttling' process involves moving an ion from one point to another along a straight trajectory, facilitated by electrodes and typically comprising split, move, and merge operations. For QCCD structures where modules are not linearly aligned, `directional turning' or `junction transport' is sometimes necessary, altering the ion's path of motion. The time cost of these additional steps is contingent on the design of the electrode structures and the connectivity channels between modules. Detailed estimations of these costs are based on the existing literature and will be shown in Section~\ref{sec:evaluation}.

\subsection{Motivation}
Before introducing the S-SYNC pipeline, we first highlight several key observations in QCCD systems that are often overlooked in previous work, such as the dynamic impact of shuttling on the topology graph and the integration of SWAP operations. 

\begin{figure}[h]
    \centering
    \includegraphics[width=0.8\linewidth]{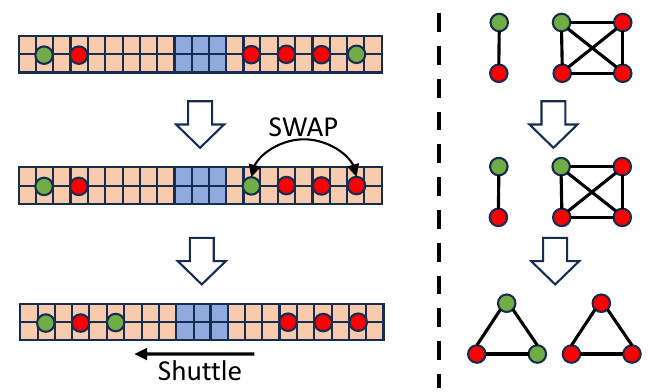}
    \caption{Illustration of Observation 1 and 2. (Left) To apply a two-qubit gate on the \textcolor{black}{two green qubits (with white space qubits omitted)}, a SWAP gate and a shuttle operation are used. (Right) The topology graph shows that, because each trap maintains full connectivity, the graph's isomorphism is preserved after the initial SWAP insertion, indicating no topological changes. However, the shuttle operation modifies the topology graph.
    }
    \label{fig:ob2}
\end{figure}

\paragraph{Observation 1 - Shuttle changes the topology graph} As illustrated in Fig.~\ref{fig:ob2}, the topology graph of a QCCD device is not always connected. This lack of connectivity prevents direct implementation of circuits using only SWAP gates. Furthermore, shuttle operations alter the positions of qubits within the QCCD, thereby modifying the topology graph dynamically. Existing compilers designed for superconducting architectures~\cite{li2019tackling, tan2020optimal, zhang2021time} are unable to accommodate these changes, highlighting the need for new methods to address the unique challenges introduced by shuttle operations.

\paragraph{Observation 2 -  Shuttle operations are typically accompanied by SWAP gates}
It is important to note that qubits can only be split at the edges of traps. When the compiler selects qubits for shuttle operations that are not positioned at these edges, additional SWAP gates must be inserted to enable the shuttle operation. These SWAP gates will increase the number of two-qubit gates, which is another key factor impact the success rate of the application.

\begin{figure}[h]
    \centering
    \includegraphics[width=0.8\linewidth]{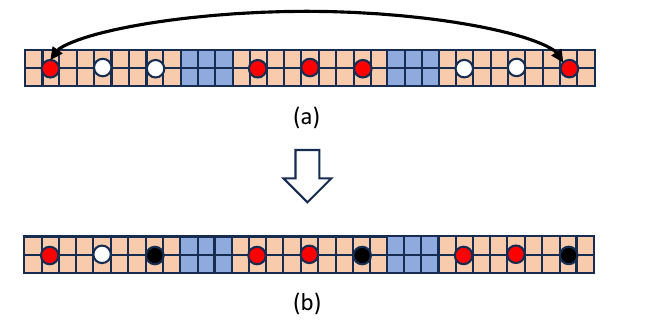}
    \caption{Illustration of Observation 3. \textcolor{black}{Red dots represent qubits, white dots denote spaces for incoming ions, and black dots indicate fixed positions for shuttling.} In the scenario shown in (a), suppose a two-qubit gate must be applied to two qubits, but the path is obstructed by a fully occupied trap. (b) To address this, a common approach in previous work~\cite{murali2020architecting} is to reserve one or two qubit positions in each trap exclusively for shuttling.
    }
    \label{fig:ob1}
\end{figure}

\paragraph{Observation 3 - Ineffectiveness of spatial utilization}
Current compilers are not optimized for spatial utilization within traps. For example, as discussed in~\cite{murali2020architecting_iontrap}, each trap typically forced to maintain two fixed free spaces to prevent blockages during qubit routing. A simple illustration of this approach is provided in Fig.~\ref{fig:ob1}. While this method prevents deadlock during scheduling, it reduces the usable space within each trap, potentially leading to inefficient use of trap capacity.

\section{S-SYNC Compiler Pipeline}\label{sec:algorithm}
In this section, we introduce the generic swap operation and develop a heuristic search framework tailored for QCCD. Figure~\ref{fig:qccd_overview} provides an overview of our S-SYNC compiler.

\subsection{Preprocessing}
\paragraph{Mapping quantum application to DAG}
A quantum circuit is a time-ordered sequence of quantum instructions and can be naturally represented by directed acyclic graph (DAG). For each gate in the given circuit, we use a node in the dependency graph to denote it. The directed edge $(g_i,g_j)$ in the graph means a dependent relation, that is $g_j$ can only be executed after $g_i$'s execution. Vertices with zero in-degree can be executed right away (if they satisfy the topology constraint), and we remove the vertex following its execution. The transformation from quantum circuit to its DAG representation can be done efficiently with 
\textcolor{black}{
$O(|\mathcal{G}|)$, where $\mathcal{G}$ is the DAG representation of given quantum circuit and $|\mathcal{G}|$ is the total number of circuit gates.
}

\begin{figure}[h]
    \centering
    \includegraphics[width=0.8\linewidth]{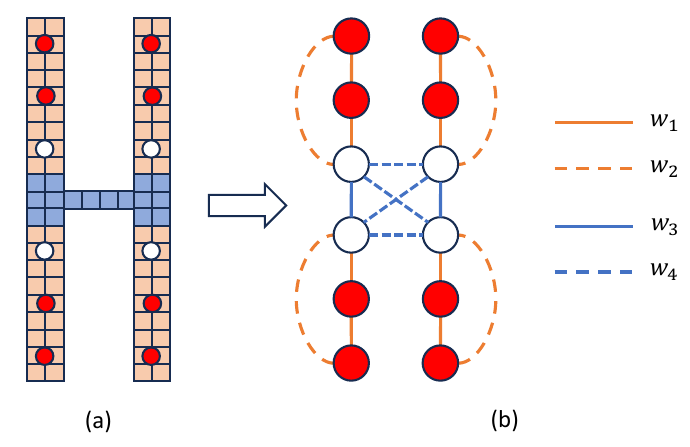}
    \caption{A demonstration of creating a graph using a given QCCD. In (a), each dot stands for a unit space; a qubit is represented by a red dot, and a free space is denoted by a white dot. The corresponding graph of (a) is shown in (b), where the blue line indicates the potential ability to swap two nodes and each orange line between two red nodes indicates that we can place a two-qubit gate on these qubits. We set edge weight and denote them by $w_i$ to characterize the communication cost varies between nodes.}
    \label{fig:QCCD formulate}
\end{figure}

\paragraph{Mapping QCCD onto a static topology model}
For the input QCCD topology, it is initially converted into the weighted connectivity graph structure, denoted as $G \in (V, E, W)$. We define each vertex $v \in V$ as a node, which can represent either a loaded qubit (red node) or an available space (white node) as shown in Fig~\ref{fig:QCCD formulate}.
The space node or available space denotes a unit of space capable of loading a single qubit. The act of moving a qubit between two traps is analogous to the interchange of these two corresponding nodes within the graph. An edge $e \in E$ connecting two nodes indicates interchangeability through operations tailored to the specific QCCD device, with its associated weight $w \in W$ represents the cost of performing the interchange. A connection between a red node and a space node signifies that the qubit is relocating to the available space, making room for incoming qubits.

The interchangeability of two nodes is typically determined by device-specific constraints in QCCD with the associated cost represented by weights, which can be shown as follows, 
\begin{enumerate}
    \item A two qubit gate act on $u,v$ can be applied iff $(u,v)\in E$ and $W(u,v)\leq threshold$, it means two qubits are placed in the same trap,
    \item To swap two nodes $(u,v) \in E$ that are both qubit nodes and have a weight $W(u,v) \leq threshold$, we require the use of a SWAP gate.
    \item With $(u,v)\in E$ and $W(u,v) > threshold$, it requires one of $u,v$ is space node. This operation is equivalent to transferring a qubit to another trap, representing the shuttle operation described earlier.
    \item If we want to swap two nodes $(u,v)\in E$ with $W(u,v) \leq threshold$ and one of them is space node, it requires these two nodes are adjacent ($W(u,v) =inner\_weight$). This operation doesn’t alter qubit arrangement but shifts spatial nodes within a trap to its edge, facilitating shuttling.
\end{enumerate}

The weights are typically defined by the cost of QCCD operations, while the threshold is determined by the cost of ion reordering. Typically, the cost of ion reordering is less than that of shuttling, which can be an indicator to represent the ion is within the trap.
Then with this formulation, particularly considering the space node, qubit interchange no longer alters the topology. By mapping the specific QCCD-device tailored operations to the graph, we could further evaluate the edge weights to identify which node changes will result in lower costs. Mapping circuit structures to weighted graphs is common, primarily for superconducting devices, with weights reflecting device parameters, such as failure rate~\cite{tannu2019not}, priority~\cite{lin2021using}, interaction frequency~\cite{bandic2023interaction}. Typically, these formulations only consider connections between qubits and do not account for space nodes.

\subsection{Generic-Swap Based Shuttling Schedule}

For the QCCD shuttling schedule task, encountering two qubits located in separate traps necessitates a method to exchange their positions by traversing the edges. Leveraging the static topology representation of QCCD devices and the DAG structure of quantum circuits, we now introduce a scheduling framework to minimize the costs associated with QCCD-specific operations, especially on minimizing the shuttle and swap operations.

It is noticed that node interchanges within the static topology graph encompass all QCCD device operations, including shuttle, move, merge, and ion reordering. For simplicity, we term this unified operation a `generic swap', as it integrates all QCCD tailored operations into the broader node interchange framework. Given that QCCD devices have the potential to scale to hundreds of qubits, obtaining an exact optimal solution becomes impractical. Consequently, we focus on heuristic algorithms to address the problem.

\begin{algorithm}
    \renewcommand{\algorithmicrequire}{\textbf{Input:}}
    \renewcommand{\algorithmicensure}{\textbf{Output:}}
    \begin{algorithmic}[1]
        \REQUIRE Circuit $\mathcal{C}$, weighted graph $G=(V,E,W)$, initial mapping $\pi$, initial space recorder $space$
        \ENSURE Gate execution list
        \STATE Build the dependency graph $\mathcal{G}$ of input circuit $\mathcal{C}$
        \STATE $wait\_list=[\,]$
        \WHILE{$\mathcal{G}$.frontier is not empty}
            \FOR{gate $g\in\mathcal{G}$.frontier}
                \IF{$g$ is executable}
                    \STATE execute $g$
                \ELSE
                    \STATE $wait\_list$.append($g$)
                \ENDIF
            \ENDFOR
            \STATE Get generic swap candidate $S(wait\_list)$
            \STATE $score=[\, ]$
            \FOR{$swap\in S$}
                \STATE $\pi_{\textcolor{black}{temp}}=\pi.update(swap)$
                \STATE $space_{\textcolor{black}{temp}} = space.update(swap)$
                \STATE $score[swap] = H(\mathcal{G}, \pi_{\textcolor{black}{temp}}, space_{\textcolor{black}{temp}})$
            \ENDFOR
            \STATE Find the $swap$ with minimal $score(swap)$
            \STATE Update $\pi$ and $space$
        \ENDWHILE
    \end{algorithmic}
\caption{Generic-Swap based Shuttling Schedule}
\label{algo:heuristic}
\end{algorithm}

The proposed heuristic algorithmic framework is outlined in Algorithm.~\ref{algo:heuristic}. The algorithm begins by checking whether the frontier of the dependency graph is empty. If so, the search algorithm ends, as all two-qubit gates in the circuit have been executed. For gates in the frontier whose qubits meet the specified requirements, the corresponding gate is executed, and the dependency graph is updated accordingly. When none of the frontier gates meet the required execution conditions, a heuristic search is employed to identify a sequence of actions that enables the execution by inserting additional operations. 

\textcolor{black}{In Step 11, the set $S$ is defined as the collection of candidate generic swaps. To construct $S$, we first examine all edges in the current graph and then include those that are valid into the set $S$.}
The $H(\cdot)$ in step 16 is the heuristic function to compute the score of each possible generic swap, which we will discuss its detail in the next subsection. For each possible generic swap, we evaluate its score as shown in Algorithm.~\ref{algo:heuristic}. Assuming that a higher score indicates a greater cost incurred during the insertion of additional QCCD device-specific operations, we select the option with the lowest score to update the mapping $\pi$. The search algorithm then continues until the frontier is empty. 

\begin{figure}[h]
    \includegraphics[width=1.0\linewidth]{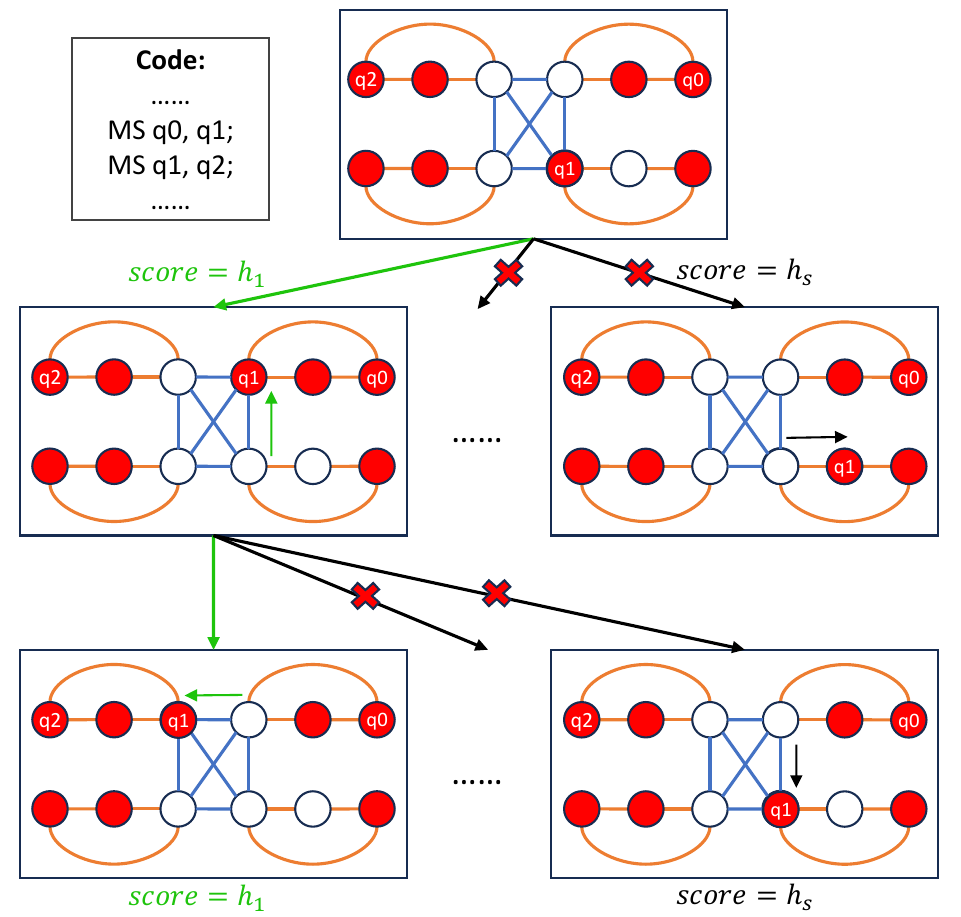}
    \caption{Illustration of the gate scheduling on QCCD  devices. Consider the scenario where a two-qubit gate is required to be applied between q0, q1 and q1, q2. The scheduling process entails aligning the qubits to ensure that corresponding qubits are positioned within the same trap prior to gate application.
    Our algorithm evaluates each path using a heuristic function and selects the one with the lowest cost (highlighted in green), as it represents the least costly operation.
    }
    \label{example of schedule}
\end{figure}

An illustrative example of employing a generic swap to schedule the circuit is depicted in Fig.~\ref{example of schedule}, wherein we suppose the necessity of applying a two-qubit gate on qubits q0 and q1, followed by q1 and q2, necessitating the use of a generic swap. 
Initially, q0 and q1 are located in different traps, requiring QCCD-specific operations to bring them into the same trap. Our heuristic algorithm explores possible operations and identifies the lowest-cost option to move q1 into the trap containing q0. The operation with the smallest cost is illustrated by the green path, representing the shuttle operation. Upon the successful execution of the two-qubit gate on q0 and q1, a similar process involving a generic swap is employed to transition q1 to the trap housing q2, thereby adhering to the operational constraints imposed by QCCD.

\subsection{Heuristic Cost Functions}
The heuristic function we used in Algorithm~\ref{algo:heuristic} step 16 is used to determine whether a generic swap is worth applying.  The heuristic function is shown below,
\begin{align}
    \label{main heuristic}
    H(swap)&=\min_g\{ decay(g) * score(g)\}+ w(swap) ,
\end{align}
 where $g\in G.frontier$.
As a penalty, we employ a decay function to discourage redundant operations.
If one of the two qubits of $g$ is involved in a generic swap recently, we have $decay(g)=1 + \delta$. Otherwise, $decay(g) = 1$.
\textcolor{black}{}
The larger the value of $\delta$, the more our algorithm is inclined to insert swap gates on new qubits, unless such a swap gate can significantly reduce the overall cost. 
The $score$ is used to evaluate the cost of executing a gate, the lower heuristic score indicates a reduced incurring costs.
\begin{align}
    score(g)&=\min\{\textcolor{black}{dis}(\pi(g.q_1)\rightarrow s_1 \rightarrow \nonumber\\ \label{gate_socre}
    &\cdots \rightarrow s_m \rightarrow \pi(g.q_2))\} + Pen(g),        
\end{align}
where $\pi(g.q_1)\rightarrow s_1 \rightarrow \cdots \rightarrow s_m \rightarrow \pi(g.q_2))$ represents the path from $\pi(g.q_1)$ to $\pi(g.q_2)$ and $s_i$ are intermediate nodes. Such a path requires some generic swap operations and has different costs based on shuttle or swap. \textcolor{black}{The term `dis' in E.q.~\ref{gate_socre} represents the cumulative weight of the path and we use $m$ to represents the maximal length of all possible paths.}
$Pen$ is a penalty function, which is the number of traps without internal space nodes. This is due to the fact that a trap cannot take part in the shuttle operation if it contains no spatial nodes, which could result in other ions being `blocked'. Therefore, it is advantageous to prioritize operations associated with lower scores, as they are likely to be more cost-effective.

\subsection{Initial Mapping}
Typically, challenges of the scheduling problem encompass two principal components: gate scheduling and initial qubit mapping. Prior research focusing on superconducting devices has underscored the profound impact of initial qubit allocation on the final quality of the circuit output~\cite{siraichi2018qubit, zulehner2018efficient, lin2023scalable, molavi2022qubit}. This indicates the necessity of devising a high-quality initial mapping to optimize circuit performance. Motivated by this insight, our investigation extends to encompass various methodologies for initial qubit mapping.

\paragraph{Two-Level Hierarchy Mapping}
We conceptualize the initial mapping process as a two-level approach. The first-level establishes the overall layout of qubits by assigning specific qubits to individual traps. The second level refines this arrangement through intra-trap mapping, adjusting the positions of qubits within each trap. The core objective of the second level is to minimize ion reordering and reduce the need for SWAP gates. Then, we first introduce three first-level mapping techniques that we consider,

\paragraph{1. Even-divided Mapping}
Due to the specific modular design of QCCD and inspiration from compilers in distributed NISQ computers, it is natural to consider mapping physical qubits evenly across each trap. We refer to this approach as uniform or even-divided mapping.

\paragraph{2. Gathering Mapping}
An alternative mapping strategy involves clustering all physical qubits as closely as possible. This approach facilitates the direct application of two-qubit gates within the same trap, thereby minimize the necessity for shuttling qubits between different traps. We designate this strategy as gathering mapping. In its implementation, we intentionally reserve a unit of space within the trap to accommodate incoming ions, ensuring that the system retains the flexibility required for subsequent qubit insertions.

\paragraph{3. STA Mapping}
STA mapping~\cite{ovide2024scaling}, a recently proposed method for linear and ring topologies in QCCD devices, builds on prior work~\cite{murali2020architecting_iontrap, saki2022muzzle} by strategically positioning qubits with stronger spatio-temporal correlations closer together.

Building on this first-level mapping, we will subsequently proceed with the optimization of the second level.
For the optimization of the second level, we will use a function $l$ to determine its location,
\begin{equation}
    l(q_i) = -\alpha E(q_i) + \beta  I(q_i)
\end{equation}
$I(q_i)$ represents the total number of two-qubit gates in the first $k$ layers of the DAG that require qubit $q_i$ to interact with other qubits located within the same trap. Conversely, $E$ denotes the total number of two-qubit gates that necessitate interactions between qubit $q_i$ and qubits residing in other traps.

To describe the arrangement of qubits within the tarp, we can utilize a queue $Q$ to represent their positions.
The elements at both ends of the sequence have the lowest scores, such that $l(Q[1]) \leq l(Q[2])\leq \cdots l(Q[k])$ for the first half and $l(Q[n]) \leq l(Q[n-1])\leq \cdots l(Q[n-k])$ for the last half, where $Q[k]$ represents the $k$-th qubit in the trap and $k$ represents the look-ahead ability and we set it to be 8 in our simulation.
As a result, the score distribution within the queue $Q$ manifests a ``mountain-like'' structure, where the scores increase towards the center and decrease towards the edges.
This is because qubits with lower scores are more likely to require shuttling to other traps, whereas qubits with higher scores are more likely to interact with the qubits within the trap. Therefore, positioning the lower-scoring qubits at the edges can reduce the overhead associated with swap gates during shuttling, while placing the higher-scoring qubits in the center can minimize the costs incurred during two-qubit gate operations.

\section{Experiment Setups}\label{sec:evaluation}

\subsection{Simulation Parameters}

\textbf{Execution Time.} Quantum gates are manipulated using laser or microwave fields in most trapped-ion platforms. \textcolor{black}{The implementation of two-qubit gates necessitates the meticulous design of corresponding pulse sequences, leading to varying time costs associated with different modulation techniques. We assume that the execution time for gates implemented using frequency modulation (FM) is proportional to the total number of ions in the chain~\cite{leung2018entangling}, and can be expressed as $\tau_{\text{FM}}(N) = \max(13.33N-54 , 100)$, where $N$ is the number of ions in the chain. The phase modulation (PM) gate~\cite{milne2020phase} is represented as $\tau_{PM}(d) = 5*d + 160$, where $d$ is the number of ions between the two ions that are being entangled. The amplitude modulation (AM) gate has two implementations, one~\cite{wu2018noise} is $\tau_{AM1}(d) = 100 * d - 22$ and another one~\cite{Trout2018AM} is $\tau_{AM2}(d) = 38 * d + 10$.}

\begin{table}[h]
\centering
 \begin{tabular}{|c |c |} 
 \hline
 Operations & Time  \\
 \hline
 Move& 5 $\mu$s   \\ 
 Split & 80 $\mu$s \\
 Merge & 80 $\mu$s \\
 Cross n-path junction & $40+20\times n$ $\mu$s \\
 \hline
\end{tabular}
\caption{List of execution times for moving, splitting, merging, and crossing junctions\textcolor{black}{~\cite{gutierrez2019transversality, blakestad2009high}}.}
\label{table:shuttling_operation_time}
\end{table}

Besides the gate implementation time, the modular nature of trapped-ion has additional costs when ions are exchanged between different modules. We then take the data from realistic experiments and set the time estimation for these extra operations in Table.~\ref{table:shuttling_operation_time}. For crossing n-junctions, we need more DACs for the steering operation, and \textcolor{black}{it can be approximated that the operation times is proportional to the channel number of junction~\cite{gutierrez2019transversality, blakestad2009high}.}

\begin{figure*}[t]
    \centering
    \includegraphics[width=0.9\linewidth]{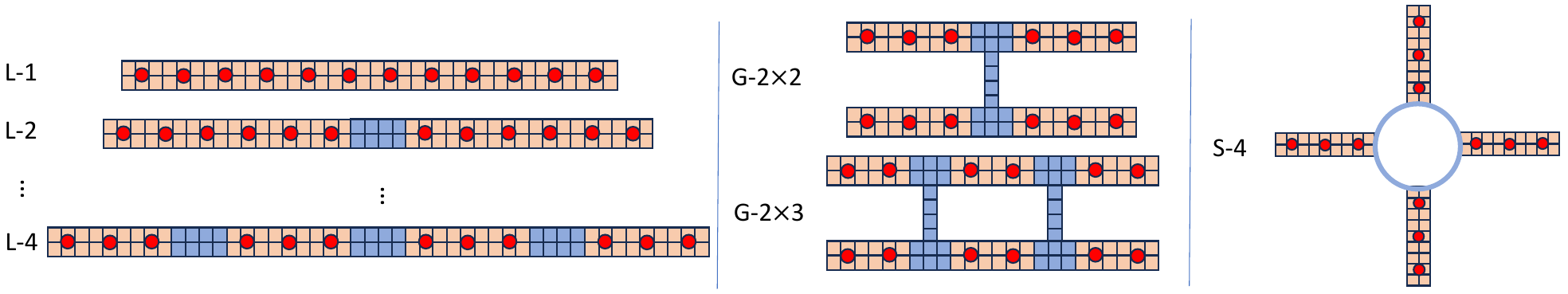}
    \caption{QCCD devices feature a variety of architectural designs, including linear-connected devices (L-series), grid-like devices (G-series), and fully-connected devices (S-series).
    }
    \label{fig:qccd_diff_archi}
\end{figure*}

\textbf{Success Rate.}
The success rate of a quantum task is a critical concern, as it is directly linked to the fidelity of each gate within the quantum circuit. In trapped-ion quantum computing, a multitude of factors can have a detrimental impact on the fidelity of quantum gates. Disturbances such as pulse amplitude and phase fluctuations, environmental thermal noise, the heating effects resulting from prolonged laser exposure on ions, and the instability of voltages and currents in the binding electrodes all significantly affect the accuracy of quantum operations. These effects become more pronounced as the number of qubits in a single trapped-ion increases, especially for two-qubit gates, where gate fidelity sharply decreases with more ions.

\textcolor{black}{The fidelity of single-qubit gates generally does not vary significantly with the number of ions, and it can achieve fidelity up to $99.9999\%$~\cite{harty2014high, smith2024single}.
The operations of separation, merging, swapping, and shuttling, required for executing two-qubit gates between different modules, involve the movement of ions. While the quantum information stored in the ions' internal states remains unchanged during spatial movement, the motional mode (phonon) used to connect the ions is strongly affected by these movements. These operations involve the acceleration and deceleration of ions through micro-electrode control, which can significantly heat the ion chain and impact the precision of subsequent two-qubit gates. Moreover, ion qubit movement operations incur a time cost. If the coherence time of the ions' internal states is short, the additional time required for these operations can reduce the overall fidelity of the quantum circuit and operations.}

Considering the energy of an ion chain is predominantly constituted by the kinetic energy of individual ions and the Coulomb potential energy between ions, the process of splitting and then merging ion chains leaves the overall Coulomb potential energy unaffected. Nevertheless, the total kinetic energy of the ion chain experiences an increment of $k_1$ quanta of motional energy, a consequence of the manipulation imposed by external electrodes. When the qubit is being shuttled from segments, it adds $k_2$ quanta of energy. \textcolor{black}{The combined effect of $k_1$ and $k_2$ leads to an increase in the total occupation number of phonon modes in the ion chain, which is typically denoted as $\bar{n}$. The product $A \bar{n}$ quantifies the overall transport impact on fidelity, where $A\propto N/\ln(N)$ is a scaling factor which represents the thermal laser beam instabilities. As the total number of ions $N$ in the chain changes with each shuttle operation, the product $A \bar{n}$ will change with every transport operation.}

Therefore, we can establish a correlation between the fidelity of two-qubit gates in trapped-ions and factors such as the number of ions in the chain, the operation time $\tau$, and \textcolor{black}{the background heating efficiency of the trap electrodes $\Gamma$. The model can treat the background heating rate as a constant. The total heat accumulation is given by $\Gamma \tau $ over different operations.} Using above analysis model, the cumulative impact of these factors on the fidelity of the quantum circuit can then be represented by,
\begin{equation}\label{eq:fidelity}
    F = 1 - \Gamma\tau - A(2\bar{n} + 1).
\end{equation}

\subsection{Benchmarks Settings}\label{subsection:benchmark_settings}

\begin{figure*}[t]
    \centering
    \includegraphics[width=0.91\linewidth]{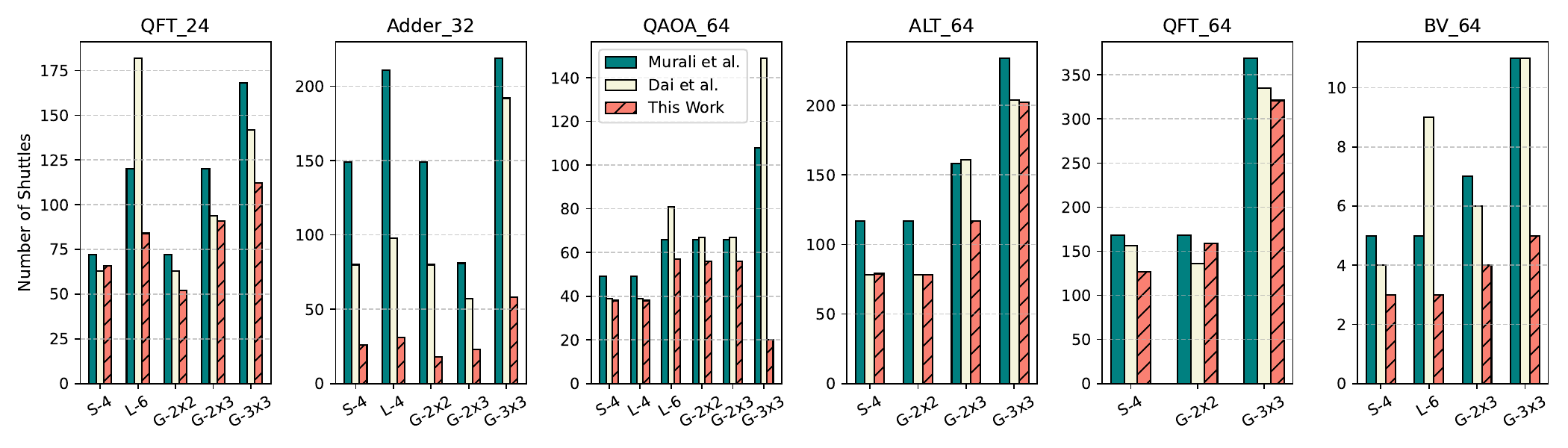}
    \caption{Comparison of shuttle counts with recent work by Murali et al.~\cite{murali2020architecting_iontrap} and Dai et al.~\cite{dai2024advanced} (Lower the better).}
    \label{fig:comparison_shuttle}
\end{figure*}

\begin{figure*}[t]
    \centering
    \includegraphics[width=0.91\linewidth]{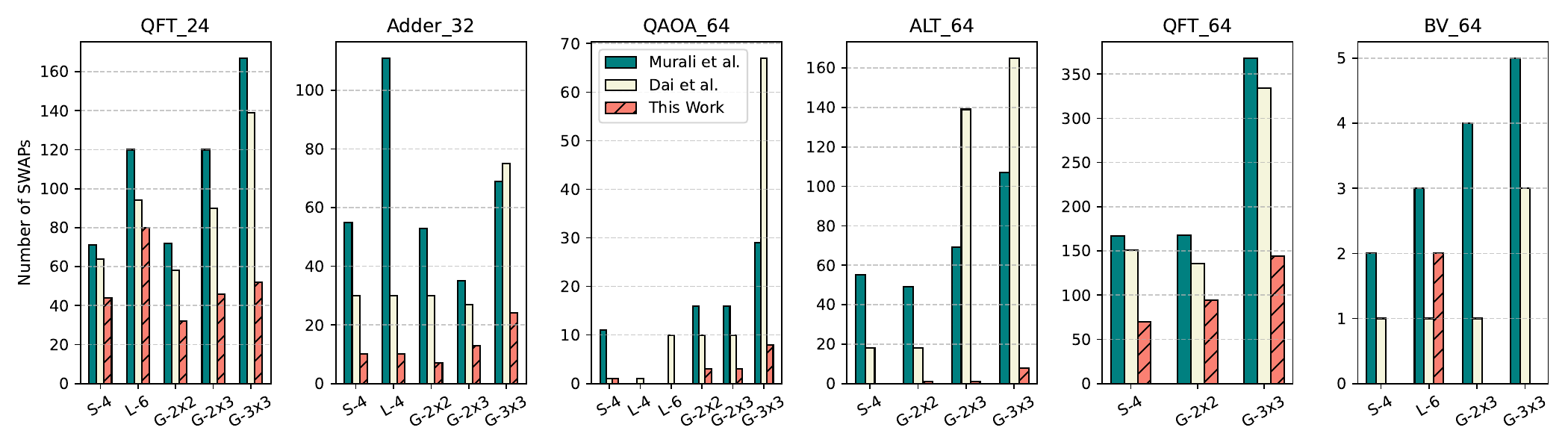}
    \caption{Comparison of SWAP gate counts with recent work by Murali et al.~\cite{murali2020architecting_iontrap} and Dai et al.~\cite{dai2024advanced} (Lower the better).}
    \label{fig:comparison_swap}
\end{figure*}

\textbf{Architectural Structures. } In order to fully leverage the robustness and performance of our algorithm on near-future trapped-ion devices\textcolor{black}{~\cite{QuanRoadmap}}, we have conducted an evaluation across various architectural structures. These structures, primarily encompass the linear-type trapped-ion and several scalable configurations as outlined in Fig~\ref{fig:qccd_diff_archi}. 
\textcolor{black}{These topology choices follow Quantinuum's hardware roadmap~\cite{QuanRoadmap}. The L-topology is analogous to ``\texttt{H2}'', while the S-topology is a variant of ``\texttt{HELIOS}''. The G-topology is designed for more advanced devices, such as ``\texttt{SOL}'' and ``\texttt{APOLL}''.} 
In detail, we select configurations S-$4$, G-2$\times$2, G-2$\times$3, and G-3$\times$3 with maximum capacities of 22, 22, 17, and 12 per trap, respectively. For certain tasks, we also test configurations L-$4$ and L-$6$, with maximum capacities of 22 and 17, respectively. The choice of maximum capacity is designed to keep the total ion count close to 200. For each device, we set the background heating rate as $\Gamma = 1$ and energy difference as $k_1 = 0.1$ and $k_2 = 0.01$, as the same as~\cite{murali2020architecting_iontrap}.

\textbf{Applications.} We evaluate our method with a set of benchmarks, as shown in Table~\ref{table:benchmarks}. These benchmarks are range from 24 to 66 qubits on QCCD. \textcolor{black}{We consider two-qubit gates and also single-qubit gates with a fidelity of $99.9999\%$}.
In detail, our benchmark suite includes selections from prior research~\cite{murali2020architecting_iontrap}, which comprises the adder circuit based on the Cuccaro adder~\cite{cuccaro2004new}, the Bernstein-Vazirani (BV) algorithm and  Quantum Approximate Optimization Algorithm (QAOA)~\cite{farhi2014quantum}. We also incorporate the QFT~\cite{bernstein1993quantum}, a critical component of Shor's algorithm~\cite{shor1994algorithms} and various phase estimation algorithms~\cite{kitaev2002classical}. Besides that, we also generate commonly used quantum machine learning ansatz such as alternating layered ansatz~\cite{Nakaji_2021}.
We source and generate the quantum program from~\cite{google2023cirq, quantum-circuit-generator, scaffCC, qiskit_contributor}.

\begin{table}[h]
\centering
\resizebox{0.92\linewidth}{!}{
 \begin{tabular}{|c | c | c | c|} 
 \hline
 \textbf{Application} & \textbf{\#Qubits} & \textbf{\#2Q Gates }& \textbf{Communication} \\
 \hline
 \multirow{1}{*}{Adder\_32} & 66 & 545 & Short-distance gates \\
 \hline
 QAOA\_64 & 64 & 1260 & Nearest-neighbor gates \\
 \hline
 ALT\_64 & 64 & 1260 &Nearest-neighbor gates \\
 \hline
 BV\_64 & 65 & 64 & \multirow{1}{*}{Long-distance gates} \\
 \hline
 \multirow{2}{*}{QFT} 
 & 24 & 552 & \multirow{2}{*}{Long-distance gates}
  \\
 & 64 & 4032 &  \\
 \hline
 Heisenberg\_48 & 48 & 13536 & Long-distance gates \\
\hline
\end{tabular}
}
\caption{List of benchmarks.}
\label{table:benchmarks}
\end{table}

\textbf{Benchmark Implementation.}
We direct use the source code from previous work by Murali et al., which is accessible in~\cite{QCCDsim}. Their qubit mapping techniques are based on a greedy heuristic algorithm adapted from~\cite{murali2019noise}. This method involves mapping the qubits into the trap and ordering the program qubits based on the sequence in which they are utilized by the application and it ensures that the trap is not fully occupied, leaving two vacant spaces only for ion shuttling. Another benchmark is used from the recent work proposed by Dai et al.~\cite{dai2024advanced}.

\textbf{Algorithm Configurations.} In our simulations, we set the decay rate as $\delta = 0.0001$. We implement a resetting mechanism for decay such that if $decay(q_i)$ has not been updated within the last five iterations, its value is reset to 1. We assign an inner weight of 0.001. The shuttle weight assigned to a segment without any junctions is $w=1$, while the weight for a path that traverses \textcolor{black}{$j$} junctions is set to \textcolor{black}{$w(j + 1)$. To illustrate, the weights are defined as follows: $w_1=0.001$ for the inner weight, $w_2=0.002$ for a distance of 2 ions, $w_3=2$ for a path crossing one junction, and $w_4=3$ for a path passing through two junctions, as shown in Fig.~\ref{fig:QCCD formulate}.}
\textcolor{black}{The initial mapping used in later evaluation is the gathering mapping.}

\textcolor{black}{The weight assignment is directly related to the shuttle cost. The more junctions traversed, the higher the weight value assigned, as each traversal through a junction causes ion heating, which subsequently decreases fidelity. Conversely, internal edges are assigned lower weight values since swap gates are less costly than shuttles. Additionally, weight are determined by the length of the qubit chain, as gates operating on qubits that are farther apart incur higher costs. 
} 

Given that the heuristic function for evaluating each potential shuttle movement may have a computational complexity of $O(n^m)$, \textcolor{black}{where $n$ is the number of qubits and $m$ is the maximal length of possible paths.} We impose a truncation limit of $m=2$ to ensure manageable computation times. \textcolor{black}{While a higher value of $m$ could improve search ability, we demonstrate that $m=2$ is sufficient to achieve near-optimal results in most cases, as shown in Fig.~\ref{fig:optimality_analysis}.}

\textbf{Simulation Platform.}
Evaluations are performed on a laptop with Intel Core i7 processor (1.4 GHz and 16GB RAM) using python 3.9.

\section{Evaluation}\label{sec:architecture_finding}

\subsection{Comparison on benchmarks}
Based on our experimental setup, we next present a comparison of the total number of shuttles, total SWAP gate count, and corresponding success rates across various benchmarks.

\textbf{Shuttle number counts.} The results are shown in Fig.~\ref{fig:comparison_shuttle}. It can show that our algorithm generally works better than~\cite{murali2020architecting_iontrap, dai2024advanced} in the number of shuttle operations. According to the application within qubit size 24, we achieve a 25\% improvement on average. For larger circuits with qubit size 64, our algorithm can also perform a more efficient shuttling compare to previous frameworks. Especially for the adder, our algorithm can achieve a reduction of up to 90.2\%.  For commonly used qml ansatz and qft, we can achieve an average reduction of 21.6\% and 10.9\% respectively.

\textbf{SWAP gate counts.} The number of SWAP comparison results are shown in Fig.~\ref{fig:comparison_swap}. It can be observed that, in most cases, the number of SWAP operations is lower compared to previous frameworks.
On average, our method achieves reductions of 68.5\% and 54.9\% compared to the frameworks in~\cite{murali2020architecting_iontrap, dai2024advanced}. Notably, for nearest-neighbor applications such as QAOA and ALT, our approach achieves SWAP reductions of up to 71.5\% and 61.7\%, respectively. We also note that in some cases, such as `BV\_64', our approach does not outperform~\cite{dai2024advanced}. However, it is still favorable as the shuttle count for this application is lower in our method. We further analyze the combined impact of these factors on the success rate below.

\begin{figure}[h]
    \centering
    \includegraphics[width=1\linewidth]{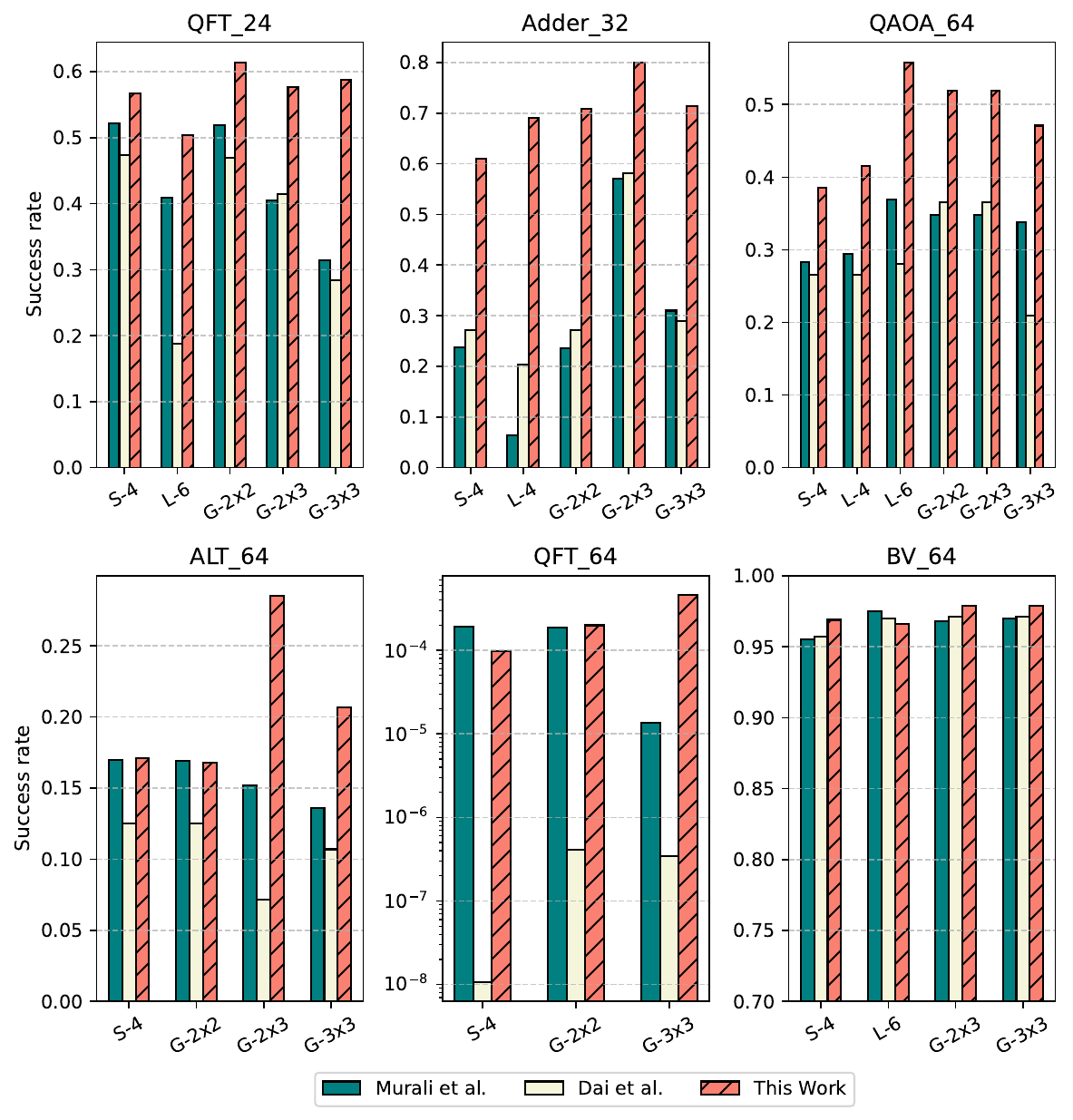}
    \caption{\textcolor{black}{Comparison of success rate with recent work by Murali et al.~\cite{murali2020architecting_iontrap} and Dai et al.~\cite{dai2024advanced} (Higher the better).}}
    \label{fig:comparison_success_rate}
\end{figure}

\textbf{Success Rate.}
As the success rate is negatively correlated to the number of shuttle operations, swap operations and the circuit execution time, we showcase the improvement in success rate achieved through our optimization achieved by implementing the FM gate.  The success rate is the product of gate fidelity, where each gate fidelity can be evaluated by Eq.~\eqref{eq:fidelity}. We show the results for different benchmarks in Fig.~\ref{fig:comparison_success_rate}. We also emphasize that some applications have a low success rate, and our focus here is to illustrate how QCCD operations can impact this outcome.

Through simulation, our algorithm can outperform the previous best-known compiler in~\cite{murali2020architecting_iontrap, dai2024advanced} on various benchmarks of different sizes on different QCCD topology. In most cases, our algorithm achieves a higher success rate, as the success rate is greatly affected by the reduction in shuttles when the number of shuttles is high. For instance, the `Adder\_32' reduces the number of shuttles by 90.2\% and its success rate has 2.3x improvement. While we have successfully optimized the number of shuttling and SWAP for benchmarks such as `QFT\_64' under specific communication topologies, it is important to note a slight decline in the success rate. 

\textcolor{black}{In most cases, shuttle and SWAP operations correlate with success rate. However, in cases like ALT, a high number of SWAP and shuttle operations can still yield a higher success rate due to the noise model. Specifically, the reduction in shuttling number may lead to a higher concentration of ions within a number of traps. This gathering effect may cause an increase in the number of ions per trap, which in turn extends the execution time of the FM gates and consequently diminishes the overall success rate. }

\begin{figure*}[t]
    \centering
    \includegraphics[width=0.9\linewidth]{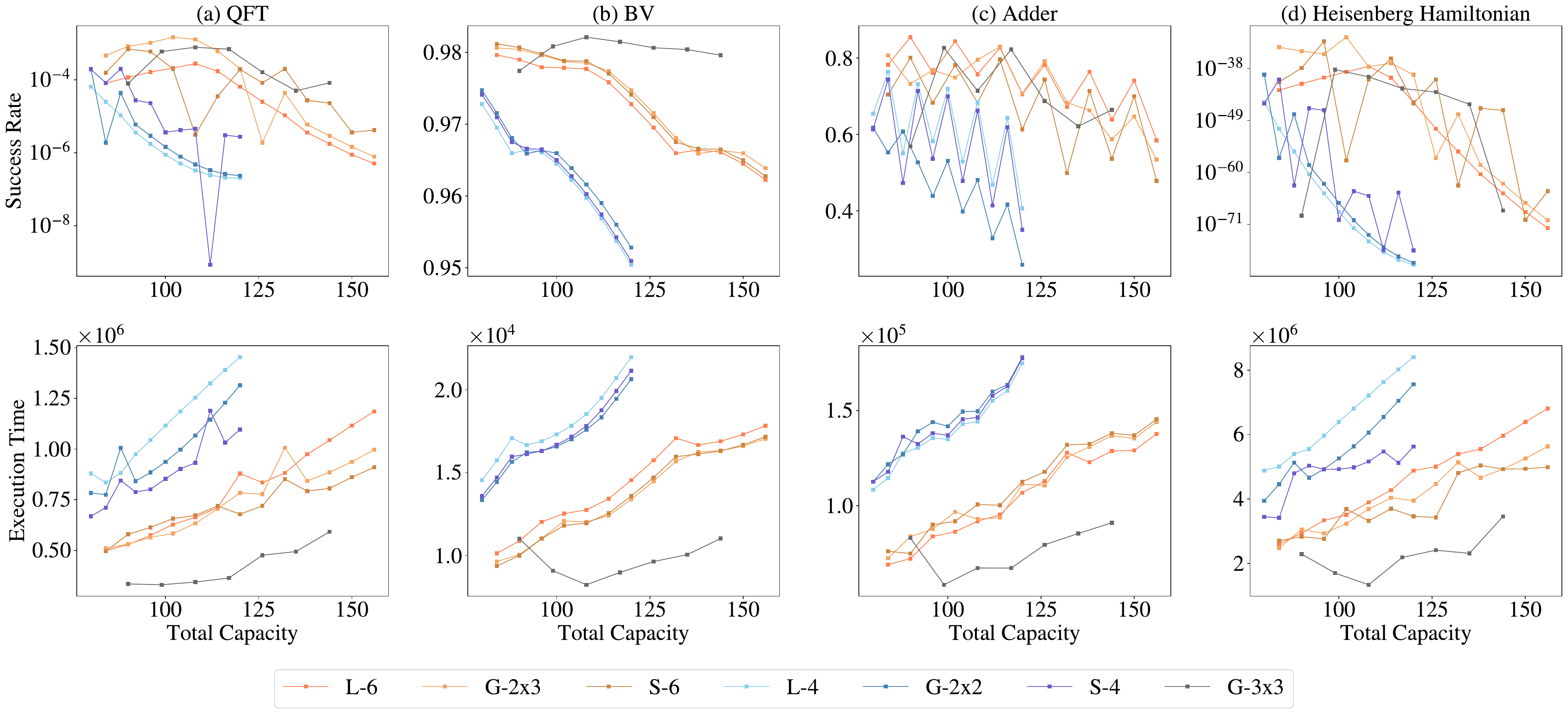}
    \caption{\textcolor{black}{Effects of communication topology and trap capacity on the fidelity of applications and their corresponding execution times. Experiments were conducted using 7 QCCD topologies to evaluate their performance across various applications, focusing on success rates and estimated execution times. The results indicate that the device topology can significantly alter outcomes.}}
    \label{fig:qccd_diff_results}
\end{figure*}

\subsection{QCCD topological analysis}\label{sec:topological_analysis}

We then assess how different topology influence quantum circuit compilation outcomes. 
Notice that the total capacity depicted in the figure represents the cumulative capacity across all systems.

\textbf{Topology.} In Fig.~\ref{fig:qccd_diff_results}, different QCCD topologies are indicated in different colors. It can be seen that the QCCD topology differs a lot on the performance. Specifically, the topology impact for BV and Adder applications is not too significant and only takes up to 1.05x and 3.6x, respectively. The topology also has a huge impact on the QFT and Heisenberg Hamiltonian simulation. Generally, the grid topology such as G-2$\times$3, G-3$\times$3 and circle topology have shown better performance in both execution time and success rate for all applications.

\textbf{Trap Capacity.} In Fig.~\ref{fig:qccd_diff_results}, we also show the execution time and success rate with respect to the total trap capacity. For topology L-6 and G-2$\times$3 topology, the trend attains a maximum and drops significantly when the trap capacity increases, which is actually fits the previous framework~\cite{murali2020architecting_iontrap} on focusing L-6 topology. Notably, among all the total trap capacity, the G-3$\times$3 topology shows the lowest execution time. The most fitted trap capacity can save up to 4x execution time reduction and can generally achieve better success rate. Empirical evidence indicates that peak performance, in terms of success rate, is typically achieved when each trap contains between 10 to 15 qubits, consistent with previous work~\cite{murali2020architecting_iontrap}.

\textcolor{black}{We note that changes in microarchitecture can affect the success rate, leading to fluctuations as trap capacity increases. These fluctuations are primarily due to qubit arrangement in applications, such as adder circuits, where certain qubits must be moved between traps based on capacity constraints.} 
Overall, the success rate of applications can be significantly influenced by topology and S-SYNC prefers grid-type topologies for better performance across diverse algorithms.

\subsection{\textcolor{black}{QCCD gate implementation analysis}}
\begin{figure}[h]
    \centering
    \includegraphics[width=0.94\linewidth]{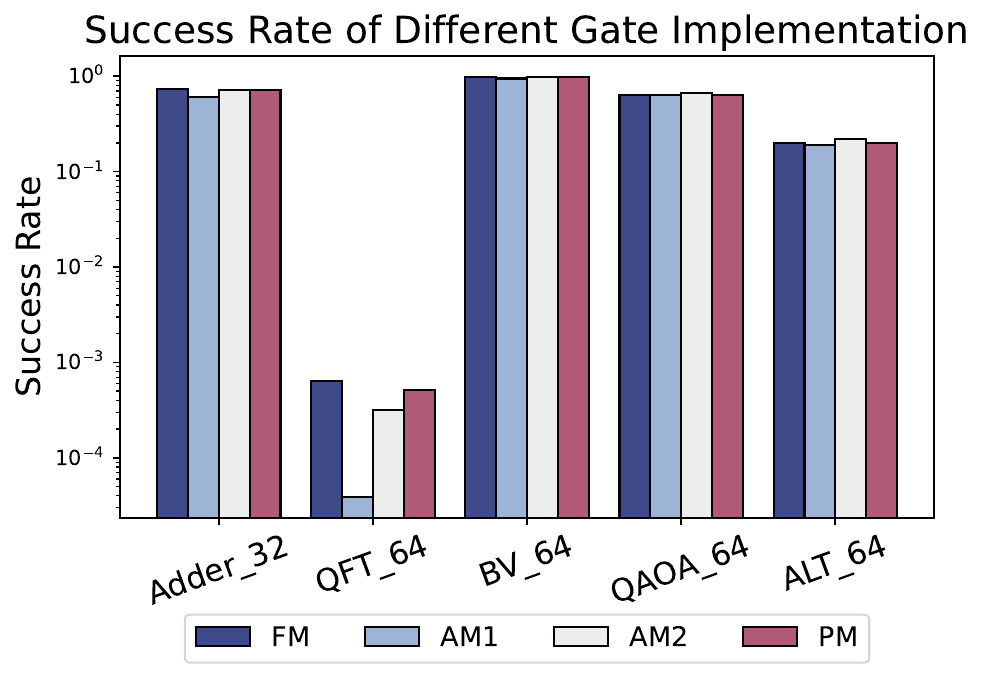}
    \caption{\textcolor{black}{Evaluation on different applications with FM, AM1, AM2 and PM gate implementations.}
    }
    \label{fig:gate_implementation}
\end{figure}

\textcolor{black}{
Furthermore, we conduct an analysis of different types of gate implementations using five large-scale applications on the G-2$\times$3 topology with a trap capacity of 16. The results are shown in Fig.\ref{fig:gate_implementation}.
}
\textcolor{black}{
For applications with short-range gates, such as QAOA and ALT, AM2 generally outperforms both PM and FM gates. In contrast, for applications requiring long-range gates, such as QFT, FM and PM are more suitable, as these gates exhibit a weaker dependence on ion separation. This suggests that AM gates are more advantageous for near-term applications, while FM and PM gates are better suited for more complex, long-range entanglement applications. More comparison of different gate implementation can be referred to~\cite{murali2020architecting_iontrap}.
}

\begin{figure*}[htbp]
    \centering
    \includegraphics[width=1\linewidth]{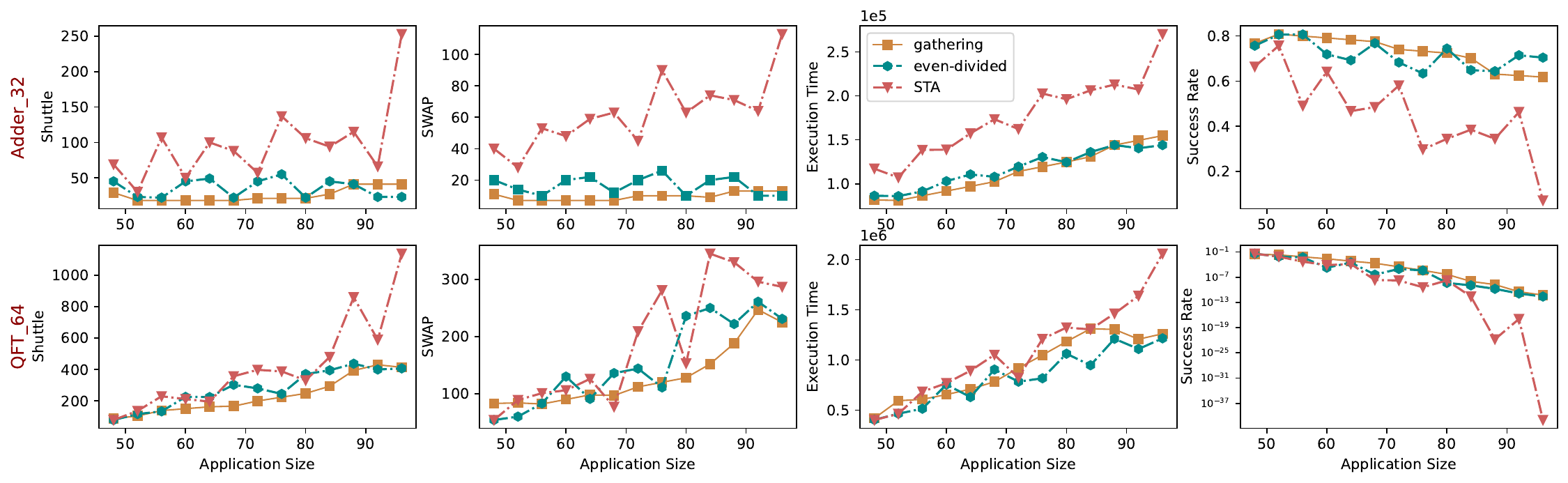}
    \caption{\textcolor{black}{Analysis of initial mapping effects. We use the 64-qubit Adder and 64-qubit QFT as examples on a G-2$\times$3 topology. The results indicate that gathering mapping results in fewer shuttling operations compared to an even-divided mapping. However, due to the nature of FM gates, this approach increases execution time, which in turn reduces the success rate.} 
    }
    \label{fig:qccd_diff_initial_mapping}
\end{figure*}

\subsection{\textcolor{black}{Comparison of Initial Mapping}}\label{subsec:shuttle_time_tradeoff}
We also study the impact of different initial qubit mapping methods. We take `Adder\_32' and `QFT\_64' as two examples to test their performance with three different mappings on G-2$\times$3 topology \textcolor{black}{with varies in application size. This setting helps prevent fluctuations as trap capacity increases and provides deeper insights into performance on S-SYNC. } In Fig.~\ref{fig:qccd_diff_initial_mapping}, we give a comparison of these initial mappings. \textcolor{black}{The results show that gathering mapping generally reduces the number of shuttles but performs poorly in terms of execution time and success rate. This lack of correlation arises from the implementation of FM gates, where the execution time is proportional to the total number of ions in the chain. The gathering mapping groups ions within the same trap to minimize shuttling, but this increases the overall execution time due to the higher number of ions in the chain, ultimately leading to a decrease in the success rate. As the complexity of an application's communication pattern increases, this phenomenon in success rate becomes more pronounced.}
This observation suggests potential for further studies on the trade-offs between different communication patterns and initial mappings.

\textcolor{black}{
\subsection{Sensitivity Analysis of Hyperparameters}
\begin{figure}[h]
    \centering
    \includegraphics[width=1\linewidth]{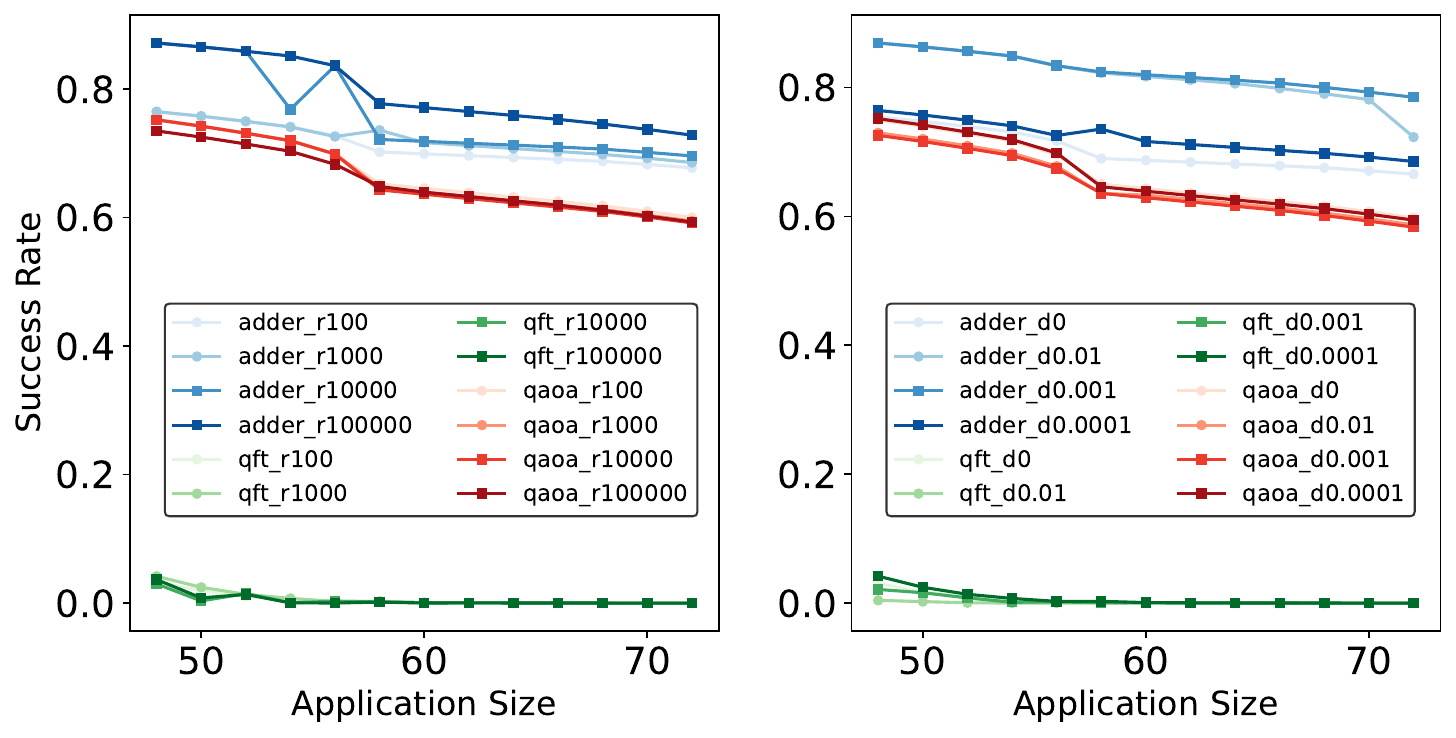}
    \caption{\textcolor{black}{Hyperparameter sensitivity analysis. (Left) Weight Analysis. r represents the ratio of shuttle weight to inner weight. (Right) Decay rate analysis. d represents the decay rate $\delta$. }}
    \label{fig:hyperparameter_sensitivity}
\end{figure}
To further analyze the impact of hyperparameters, we conduct additional simulations, as shown in Fig.~\ref{fig:hyperparameter_sensitivity}. We choose G-2$\times$2 topology with trap capacity 20.
}
\textcolor{black}{The graph weights are introduced to represent the cost associated with shuttling and SWAP operations. In the previous benchmark, we set inner weight to 0.001 and shuttle weight $w$ to 1. 
In Fig.~\ref{fig:hyperparameter_sensitivity} (left), we vary the ratio of shuttle weight to inner weight, comparing different settings ranging from r100 (ratio=100) to r100000. The results indicate that as long as the shuttle weight remains positively proportional to the inner weight, the overall performance of the algorithm remains almost consistent. In Fig.~\ref{fig:hyperparameter_sensitivity} (right), we vary the decay rate $\delta$ from 0 to 0.01.  
A higher $\delta$ prioritizes generic swaps for parallel execution, but excessive values overemphasize parallelism. Conversely, a lower $\delta$ increases the risk of local optima, leading to additional swaps. The optimal value depends on the application and architecture. In previous benchmark, we set $\delta=0.001$, which performs well across most applications.
}
\textcolor{black}{
\subsection{Scalability Analysis}
}
\begin{figure}[h]
    \centering
    \includegraphics[width=1\linewidth]{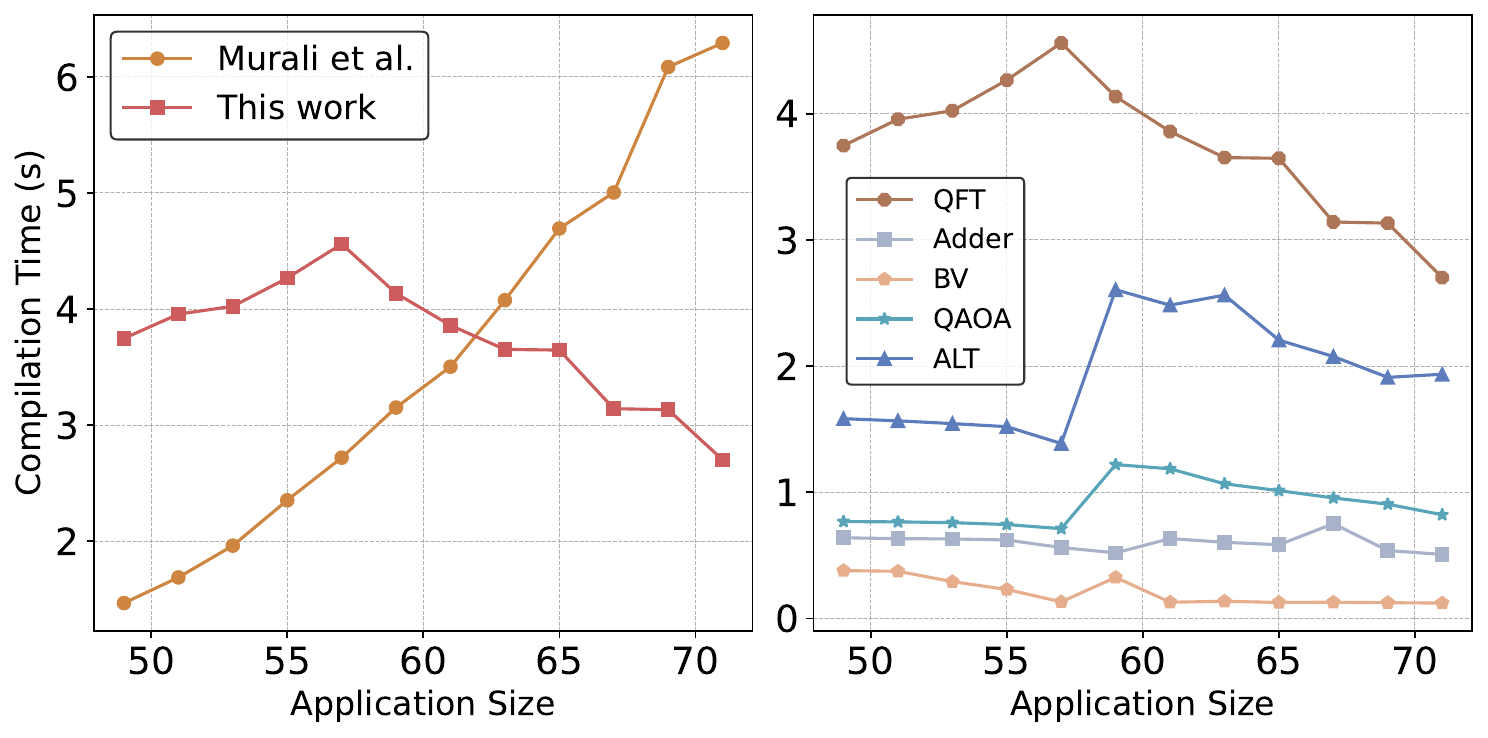}
    \caption{\textcolor{black}{Compilation time varies with the size of application.}}
    \label{fig:compilation_time}
\end{figure}
\textcolor{black}{
The compilation time is another key point in enabling executing large-scale applications~\cite{beverland2022assessing}. We also choose G-2$\times$2 topology with trap capacity 20 following the previous analysis. In Fig.~\ref{fig:compilation_time} (left), we provide a compilation time of our S-SYNC with comparison to baseline models. We select `QFT' as the example, as it is the most complex one among benchmarks. In Fig.~\ref{fig:compilation_time} (right), we benchmark the compilation time for all previously considered benchmarks.}

\textcolor{black}{The compilation time of S-SYNC does not grow approximately linearly with application size.
Initially, as the application size grows, S-SYNC's compilation time also rises. According to Eq.~\eqref{gate_socre}, this is because a larger number of space nodes creates more possible paths, increasing compilation overhead. However, as the application size continues to scale, the reduction in space nodes leads to a gradual decline in compilation time for our scheduling algorithm. This feature enhances the scalability of our approach compared to previous methods.
}

\textcolor{black}{
\subsection{Optimality Analysis}
We evaluate the optimality gap of S-SYNC by comparing it against three idealized scenarios: perfect shuttle, which assumes that all moves are fully compatible; perfect SWAP, which represents a scenario where all ions to be shuttled are positioned at the edges of the trap; and the ideal case, which combines both perfect shuttle and perfect SWAP. These scenarios establish an upper bound on the success rate through brute-force methods.
}

\begin{figure}[h]
    \centering
    \includegraphics[width=1\linewidth]{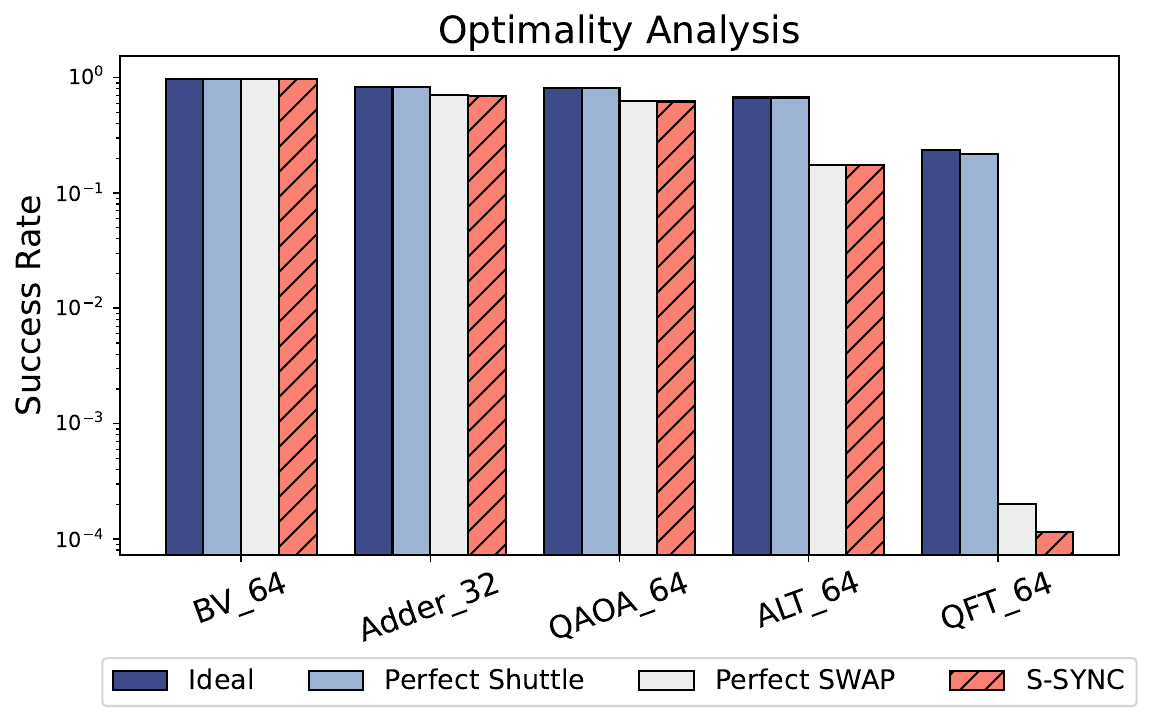}
    \caption{\textcolor{black}{Optimal Analysis of S-SYNC compared to ideal, perfect shuttle and perfect SWAP.}}
    \label{fig:optimality_analysis}
\end{figure}

\textcolor{black}{
Following the previous benchmark setting, we also adopt the G-2$\times$2 topology with a trap capacity of 20 and evaluate the performance using five benchmarks. The results in Fig.~\ref{fig:optimality_analysis}, demonstrate that for applications with simple communication patterns, S-SYNC achieves near-optimal results. However, the gap for QFT\_64 is slightly larger due to its complex communication pattern, which requires long-distance entanglement gates, increasing the number of shuttling operations and ultimately impacting the overall fidelity. Overall, S-SYNC closely matches the performance of perfect SWAP, yet an optimality gap remains when compared to perfect shuttle, highlighting the need for further research to maintain SWAP performance while improving shuttling efficiency.
}

\section{Related Works}\label{sec:related_work}

\textbf{Work with similar settings.} Previous work has primarily studied the micro-architectural design, focusing on the impact of different gate implementations and trap capacity choices on linear-type and G-2×3 QCCD devices~\cite{murali2020architecting_iontrap}. Other studies~\cite{saki2022muzzle, upadhyay2022shuttle} have focused on reducing shuttling operations in linear topologies, tailoring heuristic methods accordingly. Our approach is different from prior work by formulating the spatial configuration within the trap and introducing the generic swap to co-optimize the shuttle and swap operation requirements for all topologies. \cite{schoenberger2024using} examines an exact solution for shuttling based on Boolean satisfiability, which does not scale to the cases considered in this work. Additionally, \cite{schoenberger2024shuttling} addresses conflict-free shuttling by designating processing and memory zones for application execution, with the focus distinct from ours.

\textbf{Linear-tape trapped-ion.} There are multiple studies that focus on the optimization of shuttling operations within single-trap devices. They use methods such as integer linear programming and heuristic functions that are specifically tailored for single trapped-ion devices~\cite{wu2019ilp_iontrap, wu2021tilt_iontrap, stevens2017automating, wu2024boss}. However, scalability remains a challenge for single trapped-ion devices and the shuttling operations in these devices differs from those employed in QCCD architectures.

\textbf{Experiments carried on real device.} Number of empirical studies utilizing trapped-ion devices have been executed, demonstrating their formidable capability in executing quantum applications. Notably, Quantinuum has introduced a 32-qubit QCCD device~\cite{moses2023race}. They achieved the realization of non-Abelian topological order~\cite{iqbal2024non} and work~\cite{he2023alignment} also demonstrates its power in near-term applications. These efforts predominantly concentrate on the operational aspects of real devices, specifically those with a constrained number of qubits. In contrast, our work may serve as a step towards preparing for the forthcoming generation of QCCD devices and offers insights for architectural design.

\textbf{Future devices.} Building upon the foundation of QCCD, investigations into the challenges of connecting and compiling multi-QCCD module systems have initiated. A notable example includes the development of quantum-matter link structures, facilitated by ion--photon coupling entanglement~\cite{akhtar2023high}. Additionally, the TITAN~\cite{Chu2024titan} extended compiler scheme, which integrates ultra-fast photonic switches. These studies concentrate on the development of exceptionally large systems in the further future. Also, it has focused on reducing electrode overhead in large-scale trapped-ion systems, and~\cite{malinowski2023wire} introduces an integrated switching electronics architecture for wiring of 1000-qubit QCCD systems. These show the potential for realizing large-scale QCCD constructions in near term.

\section{Conclusion}\label{sec:conclusion}
Executing two-qubit gates on QCCD requires specific designs to reduce the shuttle and SWAP insertions associated with circuit transformation. To address this, we introduced S-SYNC to co-optimize the number of shuttling and SWAP operations. S-SYNC is equipped with an extra intermediate layer that transforms the QCCD into a weighted graph with topology inputs, considering the unification of shuttle and swap operations. Simulation results demonstrate that our algorithm surpasses the performance of existing compiler support across various topologies. For future research, it is recommended to explore various successful qubit mapping methods within our established framework~\cite{lao2021timing, zulehner2017exact, li2019tackling, tan2020optimal, zhang2021time, siraichi2019qubit, yu2023symmetrybased}. We also note that many noise-adaptive compilation policies exist for superconducting computers~\cite{murali2019noise, peters2022noise, sharma2023noise, hour2024improving, wang2022quantumnas, murali2020architecting2} to tailor applications for noisy quantum hardware. 
These methods can be adapted to our context to provide further insights and explore trade-offs regarding the noisy performance of near-future QCCD devices.

\section*{Acknowledgments}
We thank Zhixin Song and anonymous reviewers for helpful comments.
This work was partially supported by the National Key R\&D Program of China (Grant No.~2024YFE0102500),  the National Natural Science Foundation of China (Grant. No.~12447107), the Guangdong Natural Science Foundation (Grant No.~2025A1515012834), the Guangdong Provincial Quantum Science Strategic Initiative (Grant No.~GDZX2403008), the Guangdong Provincial Key Lab of Integrated Communication, Sensing and Computation for Ubiquitous Internet of Things (Grant No.~2023B1212010007), the Quantum Science Center of Guangdong-Hong Kong-Macao Greater Bay Area, and the Education Bureau of Guangzhou Municipality.

\bibliographystyle{ACM-Reference-Format}
\bibliography{sample-base}

\end{document}